\documentclass[a4paper,11pt]{article}
\pdfoutput=1 

\usepackage{jcappub} 

\usepackage[T1]{fontenc} 
\usepackage{hyperref}       
\usepackage{url}            
\usepackage{amsfonts}       
\usepackage{nicefrac}       
\usepackage{microtype}      
\usepackage{amsmath}
\usepackage{graphicx}
\usepackage{color}

\begin{document}
\title{Re-evaluating evidence for Hawking points in the CMB}

\author[a,1]{Dylan L.~Jow,\note{Corresponding author.}}
\author[b]{Douglas Scott}

\affiliation[a]{
  Canadian Institute for Theoretical Astrophysics\\
  University of Toronto\\
  60 St. George Street, Toronto, ON, Canada, M5S 3H8}
\affiliation[b]{
  Department of Physics and Astronomy\\
  University of British Columbia\\
  6224 Agricultural Road, Vancouver, BC, Canada}

\emailAdd{djow@physics.utoronto.ca}

\abstract{
We investigate recent claims for a detection of ``Hawking points'' (positions on the sky with unusually large temperature gradients between rings) in the cosmic microwave background (CMB) temperature maps at the 99.98\,\% confidence level. We find that, after marginalization over the size of the rings, an excess is detected in {\it Planck\/} satellite maps at only an 87\,\% confidence level (i.e., little more than $1\,\sigma$). Therefore, we conclude that there is no statistically significant evidence for the presence of Hawking points in the CMB.}


\notoc

\maketitle
\hfil
\flushbottom

\section{Introduction}
\label{sec:intro}

The cosmic microwave background (CMB) sky is remarkably Gaussian and statistically isotropic.
Constraints on its non-Gaussianity or breaking of statistical isotropy can be used to test models of the early Universe.   There have been many suggestions in the literature for ``anomalies'' or ``curiosities'' in the CMB data (see, e.g., Refs.~\cite{PlanckIandS2013,PlanckIandS2015,PlanckIandS2018,Schwarz2016,FrolopScott2016} for overviews).  The general idea of such studies is to carry out phenomenological searches for unusual features, with the hope that this might point towards some specific change in early Universe physics.  There is rarely a specific model being tested.  One exception is the conformal cyclic cosmology (CCC) proposed by Penrose \cite{Penrose2011,Penrose2012}, where it is asserted that there are certain features expected in the CMB.  The exact form of these features has changed since the first CCC proposal, but in the current version the prediction is for the presence of a special type of non-Gaussianity in the CMB, namely ``Hawking points'' (hereafter ``HPs''). These HPs manifest specifically as rings on the sky with a large gradient in temperature across the width of the ring, and are stated to be the results of Hawking radiation from black holes leaking into the current ``aeon'' (i.e., a distinct phase of a cyclic cosmology) from the previous aeon \cite{AnMNP2018}.  The physical mechanism for creating such regions is not clear to us, and hence it is not obvious whether the presence (or absence) of such features would argue for (or against) a cyclic cosmology.  Nevertheless, if we accept that this prediction exists, then we can at least carefully check whether or not the claimed signatures occur in the CMB with substantially higher frequency than would be expected from realizations of Gaussian skies.

An et al.\ (Ref.~\cite{AnMNP2018}) have explicitly claimed to detect an excess of HPs of a certain scale in the \textit{Planck} satellite maps of the CMB, at a confidence level of 99.98\,\%. Given the high degree with which the Gaussianity of the CMB has been tested \cite{PlanckIandS2018,PlanckNG2018}, a significant observation of HPs on the sky would have profound implications for CMB analyses and cosmology in general, and so such a claim deserves to be independently tested. In this work, we try to follow as closely as possible what was done by An et al., in order to understand whether there is really evidence for regions of the sky around which there are anomalously strong radial temperature gradients. After identifying potential HPs in the CMB temperature, we also perform stacks in temperature and polarization around the HPs to investigate whether there is any persistent pattern in the polarization corresponding to the HPs.

\section{Procedure}
\label{sec:procedure}

The procedure used here follows the methods outlined in An et al.\ \cite{AnMNP2018} and the related earlier paper by some of the same authors \cite{AnMN2018}. We search for an excess of HPs in the CMB using the 2018 full-mission and half-mission (i.e., data splits that can be used to track the noise levels) \textit{Planck} data \cite{PlanckMission2018}, specifically adopting the \texttt{SMICA} component-separation procedure \cite{PlanckCS2018}. That is, we use the CMB map and corresponding mask found in the file \texttt{COM\_CMB\_IQU-smica\_2048\_R3.00\_full.fits}, which can be downloaded from the Planck Legacy Archive.\footnote{\url{http://pla.esac.esa.int}}  We compare the data with simulated CMB temperature maps generated using \texttt{HEALPix}\footnote{See \url{http://healpix.sourceforge.net} for more details.} \cite{Gorski2005} routines from the best-fit \textit{Planck} power spectrum \cite{PlanckP2018}. The simulations are generated with a resolution of $N_\mathrm{side}=2048$ and smoothed with a $5'$ beam, to match the resolution and beam of the \texttt{SMICA} \textit{Planck} CMB maps.

\begin{figure}
    \centering
    \includegraphics[width=1\columnwidth]{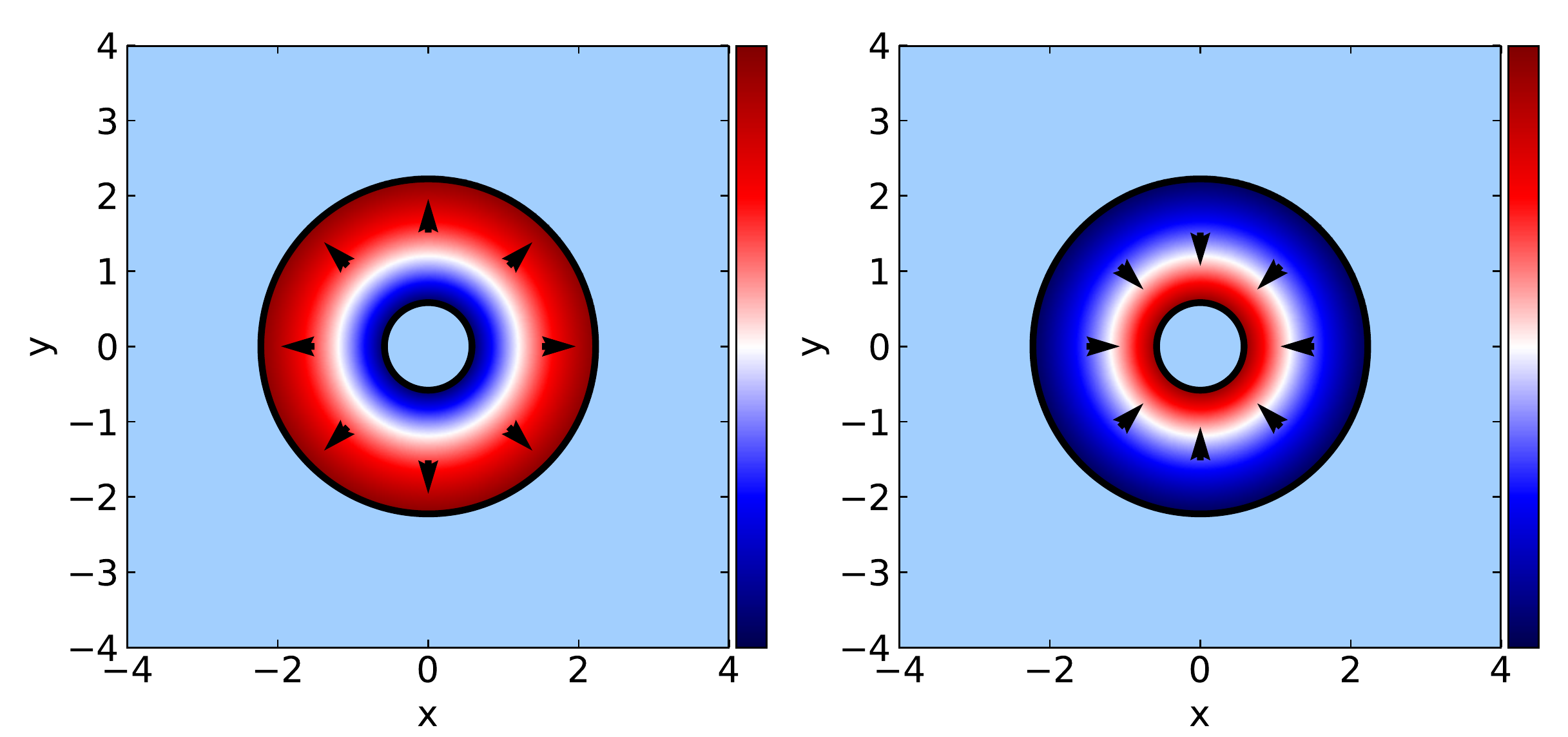}
    \caption{Schematic for the shape of the HPs in flat-sky coordinates, with $x$ and $y$ in units of degrees, and the colours representing the temperature of the CMB. The arrows show the direction of the radial gradient around the HP, where by radial we mean outwards from the centre of the HP. The left panel shows a positive HP in which the temperature increases outwards from the centre, and the right panel shows a negative HP in which the temperature increases towards the centre. Both panels show HPs for the annulus size $(r_1, \epsilon) = (0.01, 0.03) = (0.6^\circ, 1.7^\circ)$.}
    \label{fig:HPs_diag}
\end{figure}

HPs are places on the sky around which there is a strong radial temperature gradient.  More specifically, they are characterized by being the centres of rings on the sky that have a large slope in the temperature across the width of the ring
(see Figure~\ref{fig:HPs_diag}). For an annulus of inner radius $r_1$ and width $\epsilon$ centred around the direction $\hat{\textbf{n}}$ on the sky, we follow An et al.\ and estimate the gradient in temperature across the width of the ring as 
\begin{equation}
    a_{r_1,\epsilon}(\hat{\textbf{n}}) = \frac{N_{\rm pix} \sum_i (x_i T_i) - (\sum_i x_i) (\sum_i T_i)}{N_{\rm pix} \sum_i x^2_i - (\sum_i x_i)^2},
\label{eq:a}
\end{equation}
where the sums run over the (unmasked) pixels inside the annulus, $T_i$ is the temperature of pixel $i$, $x_i$ is the angular distance of pixel $i$ to the centre of the annulus, and $N_{\rm pix}$ is the total number of pixels in the annulus.

We would like to compute the value of $a_{r_1,\epsilon}(\hat{\textbf{n}})$ for a range of annulus sizes $(r_1,\epsilon)$, and a large number of points on the sky $\hat{\textbf{n}}$ for the \textit{Planck} data. To obtain a set of candidate points we simply choose the directions given by each of the 49,152 pixels in an $N_\mathrm{side}=64$ map, and reject those pixels for which a disc of radius 0.42\,radians ($24^\circ$) overlaps with the mask by more than 99\,\%. Since the sum in Eq.~(\ref{eq:a}) is only performed over unmasked pixels, this selection criterion is not strictly necessary, but it saves computation time by requiring that the calculation be performed over fewer points, and also avoids the consideration of directions that have very few unmasked pixels and would result in noisy estimates of the gradient. 

To determine whether the data have a statistically significant excess of HPs, we need to compare the distribution of the set of values $\{a_{r_1,\epsilon}(\hat{\textbf{n}})\}$ for the data to the expected distribution, assuming the standard cosmology. To achieve this, we compute a set of $a$ values for the same scales and directions for 1000 different Gaussian simulations of the CMB. We thus obtain, for each scale $(r_1,\epsilon)$, a sample of $1000 N_{\rm dir}$ values of $a$, where $N_{\rm dir}$ is the total number of directions on the sky for which we we compute $a$. Taking each scale in turn, we can in this way compute an estimate of the expected probability distribution function for the random variable $a_{r_1,\epsilon}$. We found that adding the \textit{Planck} half-difference maps as an estimate of the noise to the simulations did not alter the estimated probability distributions on $a_{r_1,\epsilon}$, and so we conclude that instrumental noise does not contribute significantly to the HP signal at the scales we are considering.

One needs to appreciate that rings of different sizes will have different numbers of pixels and hence different variances for the gradient.  To take this into account,
we define normalized gradients $\hat{a}_{r_1,\epsilon} = a_{r_1,\epsilon}/\sigma_{r_1,\epsilon}$, where $\sigma_{r_1,\epsilon}$ is the standard deviation of the un-normalized $a$ values from the simulations. From the probability distribution, we can compute the cumulative distribution function (CDF) for each scale, $F_{r_1,\epsilon}(\hat{a})$. Following Ref.~\cite{AnMN2018}, we consider the quantities 
\begin{align}
    A^+_{r_1,\epsilon} &= -\frac{b}{N_{\rm dir}} \sum_{i=1}^{N_{\rm dir}} \log(1-F_{r_1,\epsilon}(\hat{a}_{r_1,\epsilon}(\hat{\textbf{n}}_i))^b), \label{eq:Aplus}\\
    A^-_{r_1,\epsilon} &= -\frac{b}{N_{\rm dir}} \sum_{i=1}^{N_{\rm dir}} \log(1-[1-F_{r_1,\epsilon}(\hat{a}_{r_1,\epsilon}(\hat{\textbf{n}}_i))]^b),
    \label{eq:Aminus}
\end{align}
where we take $b=10{,}000$ as in Refs.~\cite{AnMNP2018} and \cite{AnMN2018}. These quantities are sensitive to an excess of points with large positive or negative values of $a$, respectively. Hence maps with more points in the tails of the distribution will have larger values of $A^+$ and $A^-$.  Note that other related quantities could be defined to assess the tails of the distributions instead of the above equations -- this is a choice.  Using Eqs.~\ref{eq:Aplus} and \ref{eq:Aminus} may not be the most obvious choice to make for a statistic quantifying the tails.  Nevertheless we follow exactly what was done by An et al.

Having computed the pair of numbers $\{A^+_{r_1,\epsilon},A^-_{r_1,\epsilon}\}$ for the data and simulations for each scale, An et al.\ count the number of simulations with larger values of these numbers compared to the data, $\{N^+_{r_1,\epsilon},N^-_{r_1,\epsilon}\}$. In this way, An et al.\ find that the 2015 \textit{Planck} data have a significant excess of Hawking points for $r_1 = 0.01\,$rad and $\epsilon = 0.02,\,0.03\,$rad, with a confidence level of 99.98\,\% (i.e., only 0.02\,\% of simulations showed more extreme gradients at these scales). This alone, however, is not sufficient evidence for an abnormal frequency of HPs on the sky, especially since there is no precise prediction coming from CCC for the scale at which we expect HPs to appear. The analysis performed in An et al.\ {\it does\/} scan over a variety of scales, and checks whether any of these scales, individually, have an excess of HPs. {\it However}, we would expect many simulated maps to have some scale at which they show an apparent excess of HPs points. So, in order to make a robust claim that the data show evidence for HPs, we must marginalize over some reasonable range of values of $(r_1,\epsilon)$.  This procedure was not done by An et al.\ -- in the language of modern particle physics analysis they failed to account for the ``look-elsewhere effect''. 

To properly marginalize over annulus scales (both radius and thickness), for each simulation we define
\begin{equation}
    A^{+/-}_i = \max\limits_{(r_1,\epsilon)} A^{+/-}_{(r_1,\epsilon),i},   
    \label{eq:Ai}
\end{equation}
where the index $i$ runs over the simulations. That is, for each simulation we find the values of $A^{+/-}$ that correspond to the scale with the most significant HP signal. This gives us a sample of $N_\mathrm{sim}=1000$ values each of $A^+$ and $A^-$, which enables us to estimate the probability distribution on these quantities for the most significant scales. Then, we can look to see where the values $A^{+/-}_\mathrm{dat}$ for the data lie on this probability distribution, in order to determine the significance of the Hawking-point signal, {\it including\/} consideration of the fact that different simulated skies might prefer rings of somewhat different radius or thickness.  We perform this marginalization for a set of $r_1 \in \{0.0, 0.05, 0.01, 0.015, 0.02, 0.025, 0.03, 0.035, 0.04 \}$, with $\epsilon \in \{0.01, 0.02, 0.03, 0.04, 0.05, 0.06, 0.07, 0.08 \}$ for each $r_1$, where the numbers here are in units of radians. We exclude annulus sizes such that $r_1 + \epsilon > 0.08$, as this is the largest scale considered by An et al. In total we marginalize over 52 scales. In principle, a full analysis would require that we marginalize over a continuous set of $(r_1, \epsilon)$, over all of the available parameter space. For the sake of reducing computational complexity, we restrict to the set of parameters stated. This set, however, is still larger than the range of scales considered by An et al., where in particular we sample the range of $r_1$ twice as finely, and we increase the range of $\epsilon$ from $[0, 0.04]$ to $[0, 0.08]$. Increasing the range of $\epsilon$ allows us to test for strong temperature gradients in discs (i.e., annuli with $r_1 = 0$) out to a radius of $0.08$. In addition to annuli, strong radial gradients across discs are included in the CCC prediction, since HPs are predicted to manifest as roughly Gaussian profiles centred on a particular point. Thus, an optimal search for HPs must include a search over a broad range of discs, as well as annuli. 

\section{Results}
\label{sec:results}

\begin{figure}
    \centering
    \includegraphics[width=1\columnwidth]{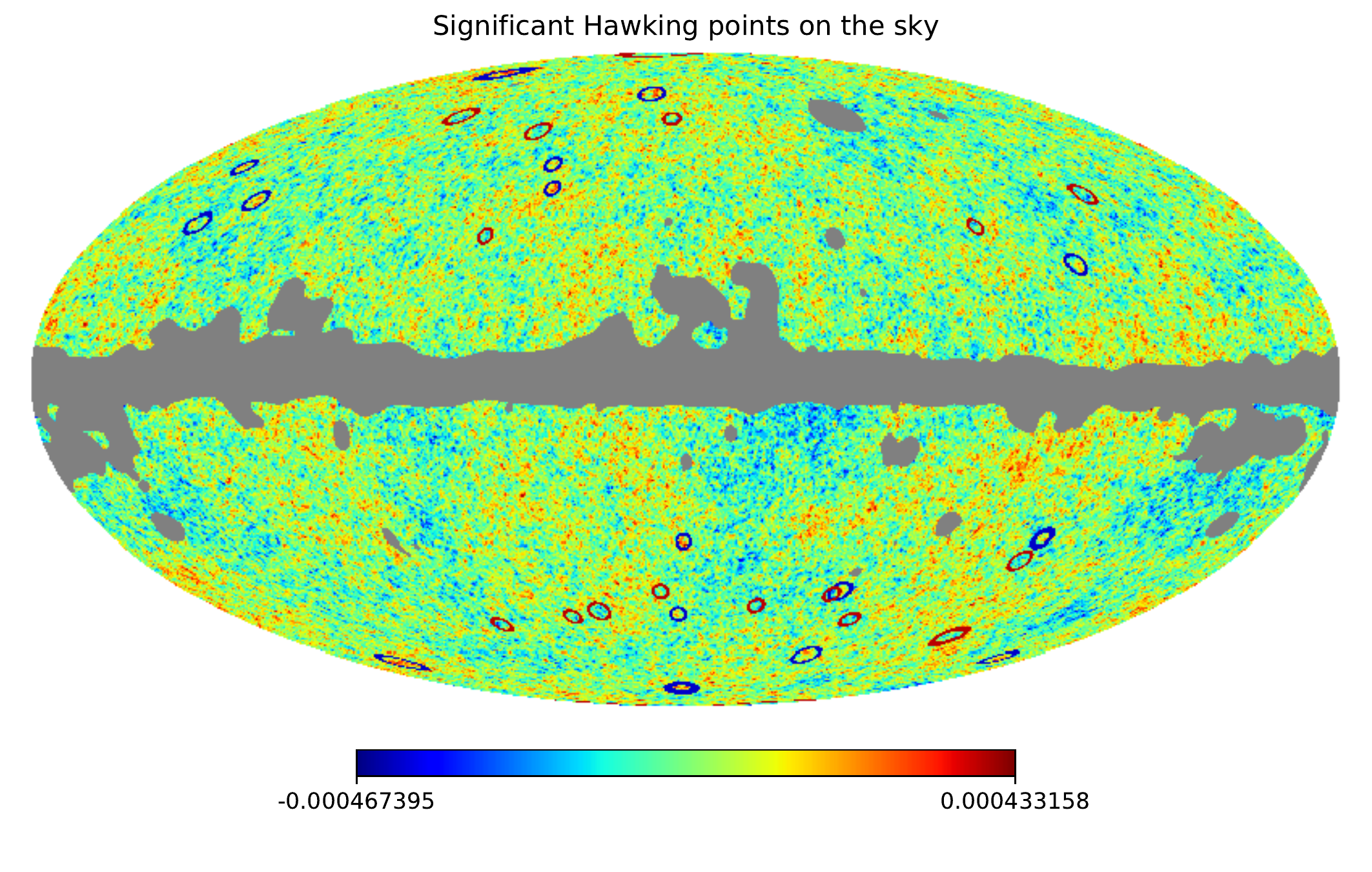}
    \caption{Mollweide projection of the sky with the most significant HPs plotted on top for the scales $(r_1, \epsilon) = (0.01, 0.02)$ and $(0.01, 0.03)$. The HPs are plotted as rings around the centre of the point, with radius $r_1+\epsilon$. The colours of the rings indicate the direction of the temperature gradient across the annulus, with red indicating outwardly increasing temperature and blue indicating decreasing temperature.}
    \label{fig:HPs_on_sky}
\end{figure}

An et al.\ \cite{AnMNP2018} claim to detect a significant excess of negative HPs (in other words, hot spots) at the scales $(r_1,\epsilon)=(0.01, 0.03)$ and $(0.01,0.02)$. Repeating the analysis performed by An et al.\ on the \textit{Planck} 2018 \texttt{SMICA} map, and comparing to 1000 simulations, we find that for these scales, $A^-_{(0.01,0.03)}=6.25$ and $A^-_{(0.01,0.02)}=8.46$, resulting in $N^-_{(0.01,0.03)}=0$ and $N^-_{(0.01,0.02)}=0$. Thus, for both of these annulus radii and widths, we observe an excess of Hawking points with a confidence of greater than 99.9\% (or in other words, a ``probability to exceed'', PTE, of less than 0.1\,\% for these scales). To get a more precise result, An et al.\ increase the number of simulations for comparison to 10,000, and find, for the scales $(r_1,\epsilon)=(0.01, 0.03)$ and $(0.01,0.02)$, a confidence of 99.98\% and 99.99\% (PTEs of 0.02\,\% and 0.01\,\%), respectively. Thus, ignoring for now the issue of marginalization, we do find results in agreement with An et al., which {\it appear\/} to show a significant excess of HPs in the data at the suggested scales. 

We note that if we recompute the values of $A^-$ for these scales, but omit the point with the highest value of $|\hat{a}|$, we obtain $A^-_{(0.01,0.02)}=5.50$ and $A^-_{(0.01,0.03)}=4.13$. Thus, about a third of the signal at these scales comes from the single most significant points. If we omit the two most significant points then $A^-_{(0.01,0.02)}=2.66$ and $A^-_{(0.01,0.03)}=2.42$, and the PTEs for these scales become greater than 10\,\%.
For $(r_1,\epsilon)=(0.01,0.02)$, An et al.\ find that the most significant point is located at $(\theta,\phi)=(2.219,0.012)$, and has a significance of $4.9\,\sigma$. For the 2018 \texttt{SMICA} data, we find that the significance of this point at the corresponding scale is $4.7\,\sigma$, and that it is only the second most significant point in the data. We find that the most significant point for $(r_1,\epsilon)=(0.01,0.02)$ is located at $(\theta,\phi)=(0.204,2.405)$ and has a significance of $4.8\,\sigma$; this corresponds to the second most significant point for this scale found by An et al. These minor disagreements between the two analyses are probably explained by the differences in the choice of specific CMB map and mask used (although it is not  made explicit in An et al.\ for which data set these values are being reported), and the fact that here we ignore masked pixels in the sum of Eq.~\ref{eq:a}, in addition to ignoring annuli that overlap substantially with the mask.

Figure~\ref{fig:HPs_on_sky} shows the full sky with the most significant HPs for the scales $(r_1, \epsilon) = (0.01, 0.02)$ and $(0.01, 0.03)$ plotted as rings. We note that the significant Hawking points at these scales do not correspond to the ``Cold Spot'' or other known larger-scale features in the temperature map described in section~6.5 of Ref.~\cite{PlanckIandS2018}.

\begin{figure}
    \centering
    \includegraphics[width=1\columnwidth]{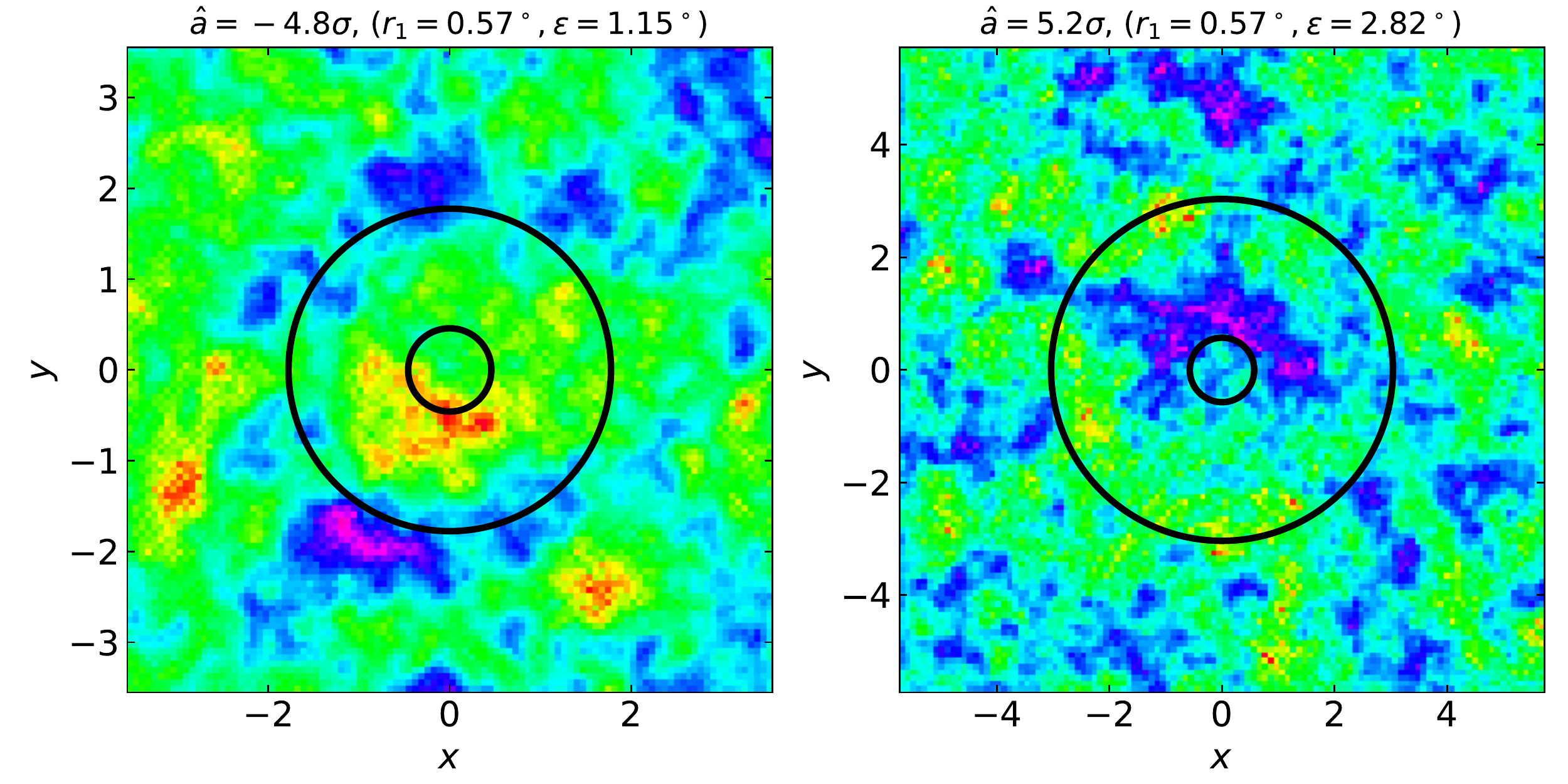}
    \caption{Examples of real and simulated Hawking points.  The left panel shows the most significant HP in the 2018 \textit{Planck} CMB map (specifically for \texttt{SMICA}). This point corresponds to an annulus with $(r_1,\epsilon)=(0.01,0.02)=(0.6^\circ,1.2^\circ)$, and has a significance of $4.8\,\sigma$ and location $(\theta,\phi)=(0.204,2.405)$, or $(l, b)=(137.8^\circ, 78.3^\circ)$ in Galactic coordinates. The point is a negative HP, meaning that the gradient of the temperature is negative going outwards from the centre. The right panel shows a positive HP of greater significance ($5.2\sigma$) found in a simulated CMB map. The black rings show the circles of radius $r_1$ and $r_1+\epsilon$, between which there is the largest radial gradient. These images are plotted using flat-sky coordinates $x$ and $y$ in units of degrees.}
    \label{fig:sig_pointl}
\end{figure}

Turning now to Figure~\ref{fig:sig_pointl}, the left panel shows the temperature profile around the most significant Hawking point on the sky. This point corresponds to an annulus with $(r_1,\epsilon)=(0.01,0.03)=(0.6^\circ,1.7^\circ)$, and has a significance of $4.8\,\sigma$ and location $(\theta,\phi)=(0.204,2.405)$, or $(l, b)=(137.8^\circ, 78.3^\circ)$ in Galactic coordinates (which is in agreement with the location of the most significant point found by An et al.\ for the same scale). As a comparison, the right panel of Figure~\ref{fig:sig_pointl} shows an example of a more significant HP found in simulated data.  We could show many other similar HPs from the simulated skies.

To assess the statistical significance of the HPs, we need to marginalize over different scales, i.e., allow each simulated sky to pick its favourite annulus radius and width, just as we did for the real sky.
As described in Section~\ref{sec:procedure}, we do this by finding the significance of the HP signal after marginalizing over $(r_1,\epsilon)$ by obtaining $A^{+/-}_i=\max\limits_{(r_1,\epsilon)} A^{+/-}_{(r_1,\epsilon),i}$ for each of the $i=1,...,1000$ simulations. We then compare the largest value for the data, $A^{-}_\mathrm{dat.} = A^-_{(0.01,0.02)}=8.46$, to the distribution of the $A^-_i$. Let $F^-$ be the cumulative distribution function that we estimate from our sample of values, $A^-_i$, from the simulations. Using this we find that $F^-(A^-_\mathrm{dat.}) = 0.87$, or a PTE of 13\,\%.  In other words, of the 1000 simulations run, 13\% have a more significant signal than the data {\it for some scale}.  We thus find that when the scales are properly marginalized over, the evidence for a statistically significant excess of Hawking points in the CMB evaporates. We performed some additional tests and found that if we were to further increase the range of scales that we perform the marginalization over, or if we increased the resolution with which we sample the scales, then we would find higher PTEs, because the extended search would increase the likelihood of finding apparent anomalies in the simulations. Since we could easily end up with a higher (and hence even less interesting) PTE value, then we consider 13\,\% to be a lower limit.  

\subsection{Clustering of HPs on the sky}
\label{sec:clustering}

In addition to the presence of HPs in the CMB, An et al.\,\cite{AnMNP2018} claim that clustering of these features could be further evidence of CCC, as one might expect the distribution of supermassive black holes in the previous aeon (the source of the observed HPs in this aeon) to be highly clustered. While we do not find a statistically significant excess of HPs in the CMB, we can still analyse the distribution of HPs on the sky.

Since the CMB has a non-trivial power spectrum, or two-point correlation function, features in the CMB such as hot spots and cold spots will not be randomly distributed with respect to each other. Even in the absence of new physics, HPs arising from statistical fluctuations in a Gaussian CMB will not be uniformly distributed on the sky, but will exhibit some amount of clustering. To test whether the distribution on the sky of the most significant HPs in the \textit{Planck} \texttt{SMICA} data differs from a purely Gaussian sky we can compare to simulations. As in Ref.~\cite{1989MNRAS.241..109S}, we use the average distance between nearest-neighbours, $\langle \theta_{\rm NN} \rangle$, as a test statistic. For each of the given scales $(r_1, \epsilon) = (0.01, 0.02)$ and $(0.01, 0.03)$ we locate the 30 most significant HPs on the sky (roughly corresponding to all the HPs with $\hat{a}>3$). For each set of HPs we compute the average distance between nearest neighbours. We find that for $(r_1, \epsilon) = (0.01, 0.02)$, $\langle \theta_{\rm NN} \rangle = 0.20\,$rad, and for $(r_1, \epsilon) = (0.01, 0.03)$, $\langle \theta_{\rm NN} \rangle = 0.19\,$rad. To determine the distribution of $\langle \theta_{\rm NN} \rangle$ for a Gaussian sky, we simulate 100 Gaussian CMB temperature maps from the \textit{Planck} best-fit power spectrum. We then compute $\langle \theta_{\rm NN} \rangle$ for the 30 most significant HPs at the corresponding scales. We find that for both $(r_1, \epsilon) = (0.01, 0.02)$ and $(0.01, 0.03)$, the distribution of $\langle \theta_{\rm NN} \rangle$ is roughly Gaussian with mean $0.20$ and $\sigma = 0.03$. Thus, for both scales, the average distance between the nearest-neighbours among the most significant HPs in the data is within 1\,$\sigma$ of the expected value for a Gaussian CMB with the \textit{Planck} best-fit power spectrum. As a result, we do not find any evidence that HPs exhibit any anomalous clustering on the sky.

\subsection{Polarization profile of HPs}
\label{sec:stack}

We have so far focused on determining whether or not there is any evidence of anomalies corresponding to a prediction of HPs in the temperature map of the CMB. However, the CMB polarization also contains information, and so it is natural to search for anomalies in the polarization as well. Without a precise prediction for the polarization profiles of HPs, it is not possible to search for specific features in the polarization that could correspond to HPs. Nevertheless, we can perform a stack of the polarization around the most significant HPs identified in the temperature map to check if there are any persistent features in the polarization in the same locations as possible HPs. 

\begin{figure}
    \centering
    \includegraphics{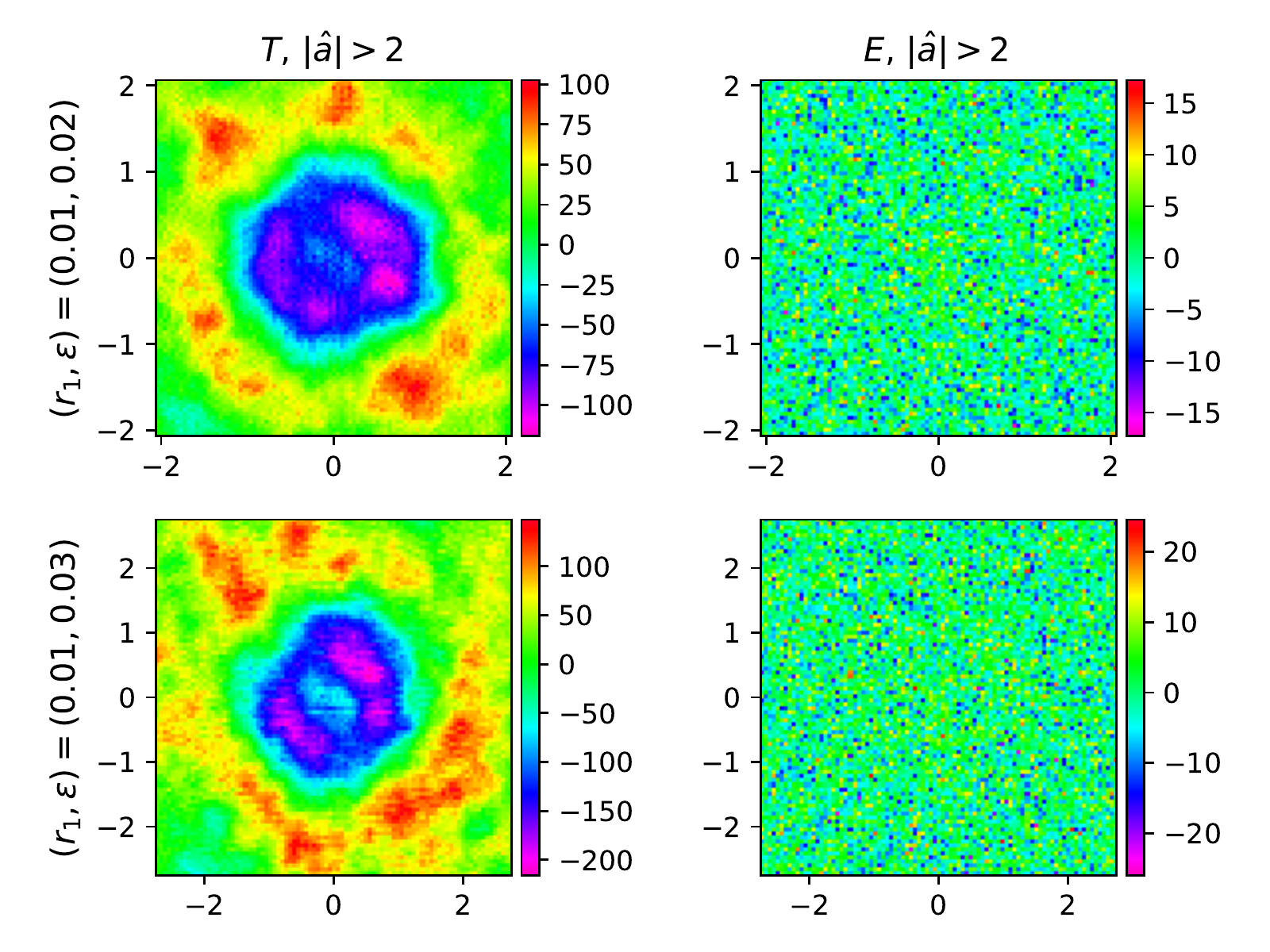}
    \caption{Stacks of the \textit{Planck} 2018 \texttt{SMICA} temperature and $E$-mode maps around HPs identified in the temperature map with a significance threshold of $2\sigma$. The left column shows the temperature stacks and the right column shows the $E$-mode stacks. The top row are stacks for HPs with $(r_1, \epsilon) = (0.01, 0.02)$ and the bottom row is for $(r_1,\epsilon) = (0.01,0.03)$. For $(r_1, \epsilon) = (0.01, 0.02)$, there were a total of 709 HPs with $|\hat{a}|>2$, and for $(r_1, \epsilon) = (0.01, 0.03)$ there was a total of 730 HPs with $|\hat{a}|>2$.}
    \label{fig:stack}
\end{figure}

To do this, we identify all HPs in the \textit{Planck} 2018 \texttt{SMICA} map that have a value of $|\hat{a}_{r_1, \epsilon}| > 2$, i.e., a 2$\sigma$ significance threshold, for the scales $(r_1, \epsilon) = (0.01, 0.02)$ and $(0.01, 0.03)$. For both the temperature and $E$-mode polarization map, we project a region around each of the identified HP locations (onto flat-sky coordinates with the $y$-axis pointing north) and then add each projection together. For HPs with $\hat{a}<0$, we add the profiles with a relative minus sign. Figure~\ref{fig:stack} shows the results of this procedure. The temperature stacks exhibit the expected ring structure of the HPs (because they were selected to have these profiles); however, there does not appear to be any corresponding structure in the $E$-mode stacks. Thus, in addition to finding no statistically significant evidence for an HP signal in the CMB temperature, we also find no evidence for a corresponding signal in polarization. 

\section{Discussion and conclusions}
\label{sec:conclusions}

The first suggestion for effects on the CMB sky from the CCC concept was for rings of low
variance \cite{Gurzadyan2010}.  These claims were quickly refuted by three independent analyses
\cite{Moss2011, Hajian2011, Wehus2011}, which showed that the original study had not properly accounted for the standard structure in the CMB sky.  There were then some counter arguments \cite{Gurzadyan2010b,Gurzadyan2011}, followed by further analysis that showed no significant results for low-variance rings \cite{Bielewicz2013}.  The specific claim then changed to being for {\it sets\/} of concentric low-variance rings \cite{Gurzadyan2013}.  This claim was also tested, and the significance of sets of low-variance rings was not confirmed \cite{DeAbreu2015}.  Now there is yet another revision of the stated CCC prediction, to high-gradient rings, rather than low-variance ones.

In this paper, we have attempted to investigate this third version of the assertion for how CCC would manifest itself.  Specifically there is the extraordinary claim that the CMB contains an excess of rings with a large gradient in temperature across them, i.e., ``Hawking points''. We found that while, for a given scale, the CMB indeed appears to have an excess of HPs, this excess ceases to be statistically significant once one marginalizes over the relevant scales. We find that Gaussian simulations of the sky contain a more significant HP signal than the data 13\,\% of the time (more if we marginalized over a wider range of scales), so that the excess is only detected at about the $1\,\sigma$ level. Therefore, the observed HP signal does not require one to appeal to exotic cosmologies, but can be adequately explained as a statistical variation. The only way this conclusion could have been avoided is if a much more dramatic set of features had been found in the first place (so that the significance would be hardly weakened by marginalization) or if the predictions of CCC could have been made much more precise before looking at the data. We also investigated whether there is a corresponding signal in the CMB polarization by stacking on potential temperature HPs in the $E$-mode data, but found no evidence of persistent features in polarization at those locations on the sky. 

\acknowledgments 
This work is based on observations obtained with Planck (\url{http://www.esa.int/Planck}), an ESA science mission with instruments and contributions directly funded by ESA Member States, NASA, and the CSA. Some of the results in this paper have been derived using the \texttt{HEALPix} package. We thank James Zibin, Dagoberto Contreras and Ren\'ee Hlo\u{z}ek for useful discussions.

\bibliographystyle{JHEP}  
\bibliography{bibliography.bib} 

\providecommand{\href}[2]{#2}\begingroup\raggedright\begin{thebibliography}{10}

\bibitem{PlanckIandS2013}
{Planck Collaboration XXIII}, \emph{{\textit{Planck} 2013 results. XXIII.
  Isotropy and statistics of the CMB}},
  \href{http://dx.doi.org/10.1051/0004-6361/201321534}{\emph{AAP} {\bf 571}
  (2014) A23}, [\href{http://arxiv.org/abs/1303.5083}{{\tt 1303.5083}}].

\bibitem{PlanckIandS2015}
{Planck Collaboration XVI}, \emph{{\textit{Planck} 2015 results. XVI. Isotropy
  and statistics of the CMB}},
  \href{http://dx.doi.org/10.1051/0004-6361/201526681}{\emph{AAP} {\bf 594}
  (2016) A16}, [\href{http://arxiv.org/abs/1506.07135}{{\tt 1506.07135}}].

\bibitem{PlanckIandS2018}
{Planck Collaboration VII}, \emph{{\textit{Planck} 2018 results. VII. Isotropy
  and statistics}}, {\emph{AAP, submitted} (2018) }.

\bibitem{Schwarz2016}
D.~J. {Schwarz}, C.~J. {Copi}, D.~{Huterer} and G.~D. {Starkman}, \emph{{CMB
  anomalies after Planck}},
  \href{http://dx.doi.org/10.1088/0264-9381/33/18/184001}{\emph{Classical and
  Quantum Gravity} {\bf 33} (Sept., 2016) 184001},
  [\href{http://arxiv.org/abs/1510.07929}{{\tt 1510.07929}}].

\bibitem{FrolopScott2016}
A.~{Frolop} and D.~{Scott}, \emph{{Pi in the sky}}, {\emph{arXiv e-prints}
  (Mar., 2016) }, [\href{http://arxiv.org/abs/1603.09703}{{\tt 1603.09703}}].

\bibitem{Penrose2011}
R.~{Penrose}, \emph{Cycles of Time: An Extraordinary New View of the Universe}.
\newblock Vintage, London, 2011.

\bibitem{Penrose2012}
R.~{Penrose}, \emph{{The basic ideas of conformal cyclic cosmology}},  in
  \emph{American Institute of Physics Conference Series} (J.~{Kouneiher},
  C.~{Barbachoux}, T.~{Masson} and D.~{Vey}, eds.), vol.~1446 of \emph{American
  Institute of Physics Conference Series}, pp.~233--243, June, 2012.
\newblock \href{http://dx.doi.org/10.1063/1.4727997}{DOI}.

\bibitem{AnMNP2018}
D.~{An}, K.~A. {Meissner}, P.~{Nurowski} and R.~{Penrose}, \emph{{Apparent
  evidence for Hawking points in the CMB Sky}}, {\emph{arXiv e-prints} (Aug,
  2018) arXiv:1808.01740}, [\href{http://arxiv.org/abs/1808.01740}{{\tt
  1808.01740}}].

\bibitem{PlanckNG2018}
{Planck Collaboration IX}, \emph{{\textit{Planck} 2018 results. IX. Constraints
  on primordial non-Gaussianity}}, {\emph{AAP, submitted} (2018) }.

\bibitem{AnMN2018}
D.~{An}, K.~A. {Meissner} and P.~{Nurowski}, \emph{{Ring-type structures in the
  Planck map of the CMB}},
  \href{http://dx.doi.org/10.1093/mnras/stx2299}{\emph{MNRAS} {\bf 473} (Jan.,
  2018) 3251--3255}.

\bibitem{PlanckMission2018}
{Planck Collaboration I}, \emph{{\textit{Planck} 2018 results. I. Overview, and
  the cosmological legacy of \textit{Planck}}}, {\emph{AAP, submitted} (2018)
  }, [\href{http://arxiv.org/abs/1807.06205}{{\tt 1807.06205}}].

\bibitem{PlanckCS2018}
{Planck Collaboration IV}, \emph{{\textit{Planck} 2018 results. IV. Diffuse
  component separation}}, {\emph{AAP, in press} (2018) },
  [\href{http://arxiv.org/abs/1807.06208}{{\tt 1807.06208}}].

\bibitem{Gorski2005}
K.~M. {G{\'o}rski}, E.~{Hivon}, A.~J. {Banday}, B.~D. {Wandelt}, F.~K.
  {Hansen}, M.~{Reinecke} et~al., \emph{{HEALPix: A Framework for
  High-Resolution Discretization and Fast Analysis of Data Distributed on the
  Sphere}}, \href{http://dx.doi.org/10.1086/427976}{\emph{APJ} {\bf 622} (Apr.,
  2005) 759--771}, [\href{http://arxiv.org/abs/astro-ph/0409513}{{\tt
  astro-ph/0409513}}].

\bibitem{PlanckP2018}
{Planck Collaboration VI}, \emph{{\textit{Planck} 2018 results. VI.
  Cosmological parameters}}, {\emph{AAP, submitted} (2018) },
  [\href{http://arxiv.org/abs/1807.06209}{{\tt 1807.06209}}].

\bibitem{1989MNRAS.241..109S}
D.~{Scott} and C.~A. {Tout}, \emph{{Nearest neighbour analysis of random
  distributions on a sphere.}},
  \href{http://dx.doi.org/10.1093/mnras/241.2.109}{\emph{MNRAS} {\bf 241} (Nov,
  1989) 109--117}.

\bibitem{Gurzadyan2010}
V.~G. {Gurzadyan} and R.~{Penrose}, \emph{{Concentric circles in WMAP data may
  provide evidence of violent pre-Big-Bang activity}}, {\emph{arXiv e-prints}
  (Nov., 2010) }, [\href{http://arxiv.org/abs/1011.3706}{{\tt 1011.3706}}].

\bibitem{Moss2011}
A.~{Moss}, D.~{Scott} and J.~P. {Zibin}, \emph{{No evidence for anomalously low
  variance circles on the sky}},
  \href{http://dx.doi.org/10.1088/1475-7516/2011/04/033}{\emph{JCAP} {\bf 4}
  (Apr., 2011) 033}, [\href{http://arxiv.org/abs/1012.1305}{{\tt 1012.1305}}].

\bibitem{Hajian2011}
A.~{Hajian}, \emph{{Are there Echoes from the Pre-big-bang Universe? A Search
  for Low-variance Circles in the Cosmic Microwave Background Sky}},
  \href{http://dx.doi.org/10.1088/0004-637X/740/2/52}{\emph{APJ} {\bf 740}
  (Oct., 2011) 52}, [\href{http://arxiv.org/abs/1012.1656}{{\tt 1012.1656}}].

\bibitem{Wehus2011}
I.~K. {Wehus} and H.~K. {Eriksen}, \emph{{A Search for Concentric Circles in
  the 7 Year Wilkinson Microwave Anisotropy Probe Temperature Sky Maps}},
  \href{http://dx.doi.org/10.1088/2041-8205/733/2/L29}{\emph{APJL} {\bf 733}
  (June, 2011) L29}, [\href{http://arxiv.org/abs/1012.1268}{{\tt 1012.1268}}].

\bibitem{Gurzadyan2010b}
V.~G. {Gurzadyan} and R.~{Penrose}, \emph{{More on the low variance circles in
  CMB sky}}, {\emph{arXiv e-prints} (Dec, 2010) arXiv:1012.1486},
  [\href{http://arxiv.org/abs/1012.1486}{{\tt 1012.1486}}].

\bibitem{Gurzadyan2011}
V.~G. {Gurzadyan} and R.~{Penrose}, \emph{{CCC-predicted low-variance circles
  in CMB sky and LCDM}}, {\emph{arXiv e-prints} (Apr, 2011) arXiv:1104.5675},
  [\href{http://arxiv.org/abs/1104.5675}{{\tt 1104.5675}}].

\bibitem{Bielewicz2013}
P.~{Bielewicz}, B.~D. {Wandelt} and A.~J. {Banday}, \emph{{A search for
  concentric rings with unusual variance in the 7-year WMAP temperature maps
  using a fast convolution approach}},
  \href{http://dx.doi.org/10.1093/mnras/sts424}{\emph{MNRAS} {\bf 429} (Feb.,
  2013) 1376--1385}, [\href{http://arxiv.org/abs/1207.6905}{{\tt 1207.6905}}].

\bibitem{Gurzadyan2013}
V.~G. {Gurzadyan} and R.~{Penrose}, \emph{{On CCC-predicted concentric
  low-variance circles in the CMB sky}},
  \href{http://dx.doi.org/10.1140/epjp/i2013-13022-4}{\emph{European Physical
  Journal Plus} {\bf 128} (Feb., 2013) 22},
  [\href{http://arxiv.org/abs/1302.5162}{{\tt 1302.5162}}].

\bibitem{DeAbreu2015}
A.~{DeAbreu}, D.~{Contreras} and D.~{Scott}, \emph{{Searching for concentric
  low variance circles in the cosmic microwave background}},
  \href{http://dx.doi.org/10.1088/1475-7516/2015/12/031}{\emph{JCAP} {\bf 12}
  (Dec., 2015) 031}, [\href{http://arxiv.org/abs/1508.05158}{{\tt
  1508.05158}}].

\end{thebibliography}\endgroup


\providecommand{\href}[2]{#2}\begingroup\raggedright\endgroup

\end{document}